\documentclass[aps,preprint,showpacs, superscriptaddress,onecolumn, floatfix]{revtex4}%
\usepackage{amsfonts}
\usepackage{amsmath}
\usepackage{amssymb}
\usepackage{graphicx}
\usepackage{subfigure}
\usepackage{color}
\def\be{\begin{equation}}
\def\ee{\end{equation}}
\def\bg{\begin{eqnarray}}
\def\en{\end{eqnarray}}

\def\ra{\rightarrow}

\begin{document}
\preprint{JLAB-THY-07-763, DAPNIA-07-194}

\title{Binding of Hypernuclei in the Latest Quark--Meson Coupling Model
\footnotetext{Notice: Authored by Jefferson Science Associates, LLC
under U.S. DOE Contract No. DE-AC05-06OR23177. The U.S. Government retains
a non-exclusive, paid-up, irrevocable, world-wide license to publish or
reproduce this manuscript for U.S. Government purposes.}}
\author{Pierre A. M. Guichon}
 \affiliation{SPhN-DAPNIA, CEA Saclay, F91191 Gif sur Yvette, France}
\author{Anthony W. Thomas}
\affiliation{Thomas Jefferson National Accelerator Facility,
12000 Jefferson Ave., Newport News, VA 23606, USA}
\affiliation{College of William and Mary, Williamsburg VA 23187, USA}
\author{Kazuo Tsushima}
\affiliation{Thomas Jefferson National Accelerator Facility,
12000 Jefferson Ave., Newport News, VA 23606, USA}
\affiliation{Departamento de Fisica Fundamental, Universidad de Salamanca,
Edificio Trilingue, Plza. Merced S/N, E-37008 Salamanca, Spain}
%

\begin{abstract}
The most recent development of the quark-meson coupling (QMC) model,
in which the effect of the mean scalar field in-medium on the hyperfine interaction is
also included self-consistently, is
used to compute the properties
of  hypernuclei. The
calculations for $\Lambda$ and $\Xi$ hypernuclei are
of comparable quality to earlier QMC results
without the additional parameter
needed there. Even more significantly,
the additional repulsion
associated with the increased hyperfine interaction in-medium completely changes the predictions
for $\Sigma$ hypernuclei. Whereas
in the earlier work they were bound by an amount similar to $\Lambda$ hypernuclei, here they
are unbound, in qualitative agreement
with the experimental absence of such states.
The equivalent non-relativistic potential felt by
the $\Sigma$ is repulsive inside the nuclear interior
and weakly attractive in the nuclear surface, as
suggested by the analysis of $\Sigma$-atoms.
\end{abstract}

\date{\today}
\maketitle

\section{Introduction}

The study of $\Lambda$ hypernuclei has a long and impressive history, with
the shell structure mapped out across the periodic
table~\cite{Dalitz:1978hw,Povh:1987pt,Yamamoto:1994tc,Gal:2004cf,Hashimoto:2006aw}. It is known that
the single particle spin-orbit force is
very small~\cite{Ajimura:2001na} and systematic studies of the
energy levels of light $\Lambda$ hypernuclei have enabled the extraction of
considerable detail concerning the effective $\Lambda$-N interaction. Current
studies of electroproduction at JLab will provide important new information
in this area~\cite{Hashimoto:2006aw,JLab-Expt}.

When one turns to $\Sigma$ and $\Xi$ hypernuclei,
the situation is quite
different. The special case of $^4$He aside, there is no experimental evidence for any $\Sigma$ hypernuclei~\cite{Bart:1999uh,Noumi:2001tx,Saha:2004ha},
despite extensive searches. Indeed, it seems
likely that the $\Sigma$-nucleus interaction is somewhat repulsive and that
there are no bound $\Sigma$ hypernuclei beyond A=4. In the case of the $\Xi$,
the experimental situation is very challenging, with just a handful of
observations of doubly strange nuclei. While we eagerly await studies of
$\Xi$ hypernuclei with new facilities at J-PARC and GSI-FAIR, it seems likely
that these hypernuclei may exist~\cite{Fukuda:1998bi,Khaustov:1999bz}
and it would be helpful to have a range of
predictions from various theoretical models.

There is currently considerable interest in the equation of state of dense
nuclear matter, especially in connection with the calculation of neutron star
properties~\cite{Weber:2004kj,SchaffnerBielich:2004ch}.
The density at which hyperons appear and particularly their order,
plays a crucial role in the equation of state
(EoS) -- for example, the inclusion
of hyperons in the QMC model seems essential if
the direct URCA process is to play
a role in cooling for observed neutron
stars~\cite{RikovskaStone:2006ta}. For matter
in $\beta$-equilibrium,
negatively charged hyperons are favoured
a priori and in
many models $\Sigma^-$
hyperons enter soon after the $\Lambda$.
Other models suggest that the $\Xi^-$
should appear after the $\Lambda$ (followed by the $\Xi^0$),
with the $\Sigma$
playing no role. Whichever scenario is ultimately correct, hyperon physics will
play a critical role at 2 -- 3 $\rho_0$ and a
relativistic treatment is essential~\cite{Guichon:2006er}.

In this work we apply the latest development of the quark-meson coupling (QMC)
model~\cite{RikovskaStone:2006ta}
to calculate the properties of hypernuclei. The major improvement in the
model that we use here is the inclusion of the effect of the medium on the
hyperfine interaction. This has the effect of increasing the splitting between
the $\Lambda$ and $\Sigma$ masses as the density rises. This is the prime
reason why we find that $\Sigma$ hypernuclei are unbound. On the other hand,
the model produces quite good binding energies for $\Lambda$ hypernuclei and
predicts modest binding of $\Xi$ hypernuclei.

\section{Recent improvement in the QMC model}

The QMC model was
created to provide insight into the structure of nuclear matter,
starting at the quark level~\cite{Guichon:1987jp,Guichon:1995ue,Saito:1994kg}.
Nucleon internal structure was
modeled using the MIT bag,
while the binding was described by the self-consistent coupling of the
confined quarks to $\sigma$ and $\omega$ meson fields generated by the confined quarks
in the other ``nucleons'' in the medium.
While the use of effective scalar and vector fields
to carry the forces is very similar to QHD,
the explicit treatment of the N internal structure
represents an important departure from
that classical approach. The
self-consistent response of the bound quarks to the mean $\sigma$ field leads
to a novel saturation mechanism for nuclear matter, with the enhancement of
the lower components of the valence Dirac wave functions, as the density
increases, reducing the effective $\sigma$-N coupling. This effect is naturally
represented by a single new parameter, $d$, the scalar polarizability, with the
nucleon effective mass taking the form:
\be
M_N^* = M_N - g_{\sigma N} \sigma +\frac{d}{2} (g_{\sigma N} \sigma)^2 \, .
\label{eq:effmass}
\ee
The QMC model has been used to study the binding
of $\omega$, $\eta$, $\eta^\prime$ and $D$
nuclei~\cite{Tsushima:1998qw,Bass:2005hn,Tsushima:1998ru},
as well as the effect of the medium on $K^{\pm}$
and $J/\Psi$ production~\cite{Saito:2005rv}.
It leads to a very natural explanation of the
small spin orbit force in $\Lambda$
hypernuclei~\cite{Tsushima:1997cu,Tsushima:1997rd}.
A recent extension of the same physical ideas to a
confining version of the NJL
model~\cite{Bentz:2001vc} has led to a quantitative description of
the EMC effect across the periodic
table, with a remarkable prediction of a sizeable
enhancement of the nuclear modification of
spin structure functions~\cite{Cloet:2005rt,Cloet:2006bq}.

For the present purpose, the most significant recent
development is the inclusion of
the self-consistent effect of the mean scalar field
on the familiar one-gluon-exchange
hyperfine interaction that in free space leads to
the N-$\Delta$ and $\Sigma$-$\Lambda$
mass splitting~\cite{RikovskaStone:2006ta}.
The original QMC model treated this term
as a small correction and
ignored its medium modification. However,
the term ``hyperfine'' is misleading, as the
N-$\Delta$ splitting is 30\% of the nucleon mass.
One-gluon-exchange in the bag model
is essentially determined by the magnetic moments of
the confined quarks~\cite{Chodos:1974pn}  and, in first approximation,
the hyperfine splitting is
\begin{equation}
\delta_N=\delta_\Lambda=-3 C\mu_0^2,\ \delta_\Delta=+3 C\mu_0^2,\ \delta_\Sigma=+C\mu_0(\mu_0-4 \mu_s)
\end{equation}
where $C$ is a constant, essentially independent of the
quark masses, and $\mu_0,\mu_s$ are, respectively, the magnetic moments of the confined light and strange quarks. From the experimental value in free space $(\delta_\Sigma -\delta_\Lambda)/(\delta_\Delta -\delta_N)=80/300$ we get $\mu_s/\mu_0 \sim 0.6$. This decrease  of
the magnetic moment with the increase of the
quark mass is a relativistic effect similar in
nature to the saturation mechanism of the QMC
model. In the  medium the light quark interacts
with the nuclear $\sigma$ field and this amounts
to decreasing its mass.  This in turn reduces its
eigenenergy (which is roughly the equivalent of the
constituent quark mass) and as a result
the magnetic moment of the light quarks
in medium changes
to $(1+\epsilon)\mu_0$ with $\epsilon >0$.
An  explicit calculation predicts $\epsilon\sim 7\%$
at nuclear matter density. Assuming that the
strange quark is not coupled to the nuclear
field  one thus finds that the relative change of
$\delta_\Sigma -\delta_\Lambda$, because of  the $\sigma$ field,  is
\begin{equation}
\frac{(\delta_\Sigma -\delta_\Lambda)_\sigma }{(\delta_\Sigma -\delta_\Lambda)}=
\frac{(1+\epsilon)[\mu_0(1+\epsilon) -\mu_s]}{\mu_0-\mu_s}\sim 1+3.5\epsilon \, ,
\end{equation}
where we have used $\mu_s/\mu_0 \sim 0.6$. This rough calculation
thus implies that the $\Sigma$ will be less bound than the
$\Lambda $ by $3.5\epsilon (\delta_\Sigma -\delta_\Lambda)\sim 20\rm MeV$,
which is consistent with the exact calculation described below.
Our motivation for applying the model
to finite hypernuclei is that this effect has the
right sign to help solve the major problem found
in the original QMC model,
namely that $\Sigma$ hypernuclei were not
much less bound than $\Lambda$ hypernuclei.

\begin{table}
\begin{center}\begin{tabular}{|c|c|c|c|c||c|c|c|}
\hline 
 & $m_{s}$ & $\Lambda$ & $\Sigma$ & $\Xi$ & $\Sigma^{*}$ & $\Xi^{*}$ & $\Omega$\tabularnewline
\hline 
$F_{s}=1$ & 0.341 & 1.135 & 1.176 & 1.355 & 1.416 & 1.599 & 1.784\tabularnewline
\hline 
$F_{s}=0.726$ & 0.297 & 1.107 & 1.189 & 1.325 & 1.368 & 1.507 & 1.654\tabularnewline
\hline 
Exp. &  & 1.116 & 1.195 & 1.315 & 1.385 & 1.533 & 1.672\tabularnewline
\hline
\end{tabular}\end{center}

\caption{Octet and decuplet masses in GeV. 
The  parameter $F_{s}$ and the
strange mass $m_s$ are fitted to the $\Lambda,\Sigma,\Xi$ masses. 
The last three columns are predictions for the decuplet masses.
\label{newbag}}

\end{table}

In view of the physical importance of the hyperfine
interaction, as we just explained, we decided
that it would be worthwhile to improve on the MIT bag model,
which traditionally yields only 50 MeV
of the $\Sigma-\Lambda$ mass
difference, instead of the experimental value of
80 MeV. The change which we introduce is motivated by the work
of Barnes~\cite{Barnes:1984pw}, who explained the
relatively large value of the effective strong
coupling constant, $\alpha_s$, required in the naive bag model.
That model ignored the enhancement of the relative q-q wave function at
short distance caused by the attractive color Coulomb force. The short
distance enhancement of the relative wave function is expected to be less
effective for heavier quarks.
As a phenomenological way of implementing these effects,
we multiply the hyperfine
interaction in the ususal MIT bag model
by a factor $F_s^n$, with $n$ the
number of strange quarks participating. We fix the bag model parameters 
in two steps. First, the masses of the nucleon and $\Delta$,  
together with their stability conditions, are used to determine the vacuum 
pressure, $B$, zero point energy, $Z_0$, and the strong coupling, 
$\alpha_s$, for a given value of the nucleon radius, which is set to 0.8fm 
in practice. This step is independent of the quark mass and $F_s$. 
The nucleon and $\Delta$ masses are exactly reproduced with 
$B=0.5541$ fm$^{-1}$, $Z_0=2.6422$ and $\alpha_s=0.4477$. 
Then the strange mass, $m_s$, and the parameter $F_s$ are chosen to give 
the best fit for  the $\Lambda, \, \Sigma$ and $\Xi$ masses. 
As shown  in Table~\ref{newbag}, where we also display the masses 
corresponding to $F_s=1$, we obtain an excellent fit with $F_s = 0.726$ 
and $m_s = 0.297\ GeV$. In particular, the $\Lambda-\Sigma $ splitting 
is well reproduced.  For completeness, we show in the last 3 columns 
of Table \ref{newbag} the good agreement obtained for the predicted 
masses for the rest of the baryon decuplet ($\Sigma^*, \, \Xi^*$ and $\Omega$).

\section{Application to the energy levels of hypernuclei}

In order to calculate the properties of finite hypernuclei, we construct
a simple shell model, with the nucleon core calculated
in a combination of self-consistent mean scalar and vector mean fields.
The determination of the scalar coupling to the octet baryons requires a
sophisticated self-consistency calculation. However, within the
Born-Oppenheimer approximation, which requires that the internal
structure of the nucleon has time to adjust to the local environment
(an approximation estimated in
Ref.~\cite{Guichon:1995ue} to be good at the level
of 3\% or better in finite nuclei), the result can be
parametrized in a very
practical form~\cite{RikovskaStone:2006ta}:
\begin{eqnarray}
M_{N}(\sigma) & = & M_{N}
- g_{\sigma}\sigma\nonumber \\
 &  & +\left[0.002143+0.10562R_{N}^{free}
-0.01791\left(R_{N}^{free}\right)^{2}\right]
\left(g_{\sigma}\sigma\right)^{2}
\label{eq:A18}\\
M_{\Lambda}(\sigma) & = & M_{\Lambda}-\left[0.6672+0.04638R_{N}^{free}-
0.0022\left(R_{N}^{free}\right)^{2}\right]
g_{\sigma}\sigma\nonumber \\
 &  & +\left[0.00146+0.0691R_{N}^{free}-0.00862
\left(R_{N}^{free}\right)^{2}\right]
\left(g_{\sigma}\sigma\right)^{2}
\label{eq:A20}\\
M_{\Sigma}(\sigma) & = & M_{\Sigma}-\left[0.6653-0.08244R_{N}^{free}+
0.00193\left(R_{N}^{free}\right)^{2}\right]
g_{\sigma}\sigma\nonumber \\
 &  & +\left[0.00064+0.07869R_{N}^{free}-0.0179
\left(R_{N}^{free}\right)^{2}\right]
\left(g_{\sigma}\sigma\right)^{2}\label{eq:A21}\\
M_{\Xi}(\sigma) & = & M_{\Xi}-\left[0.3331+0.00985R_{N}^{free}-
0.00287\left(R_{N}^{free}\right)^{2}\right]
g_{\sigma}\sigma\nonumber \\
 &  & +\left[-0.00032+0.0388R_{N}^{free}-
0.0054\left(R_{N}^{free}\right)^{2}\right]
\left(g_{\sigma}\sigma\right)^{2}
\, .
\label{eq:A22}
\end{eqnarray}
We take the bag radius, $R_N^{free}$, to
be 0.8fm but note that
(c.f. Fig.~1 of Ref.~\cite{RikovskaStone:2006ta})
the results are quite insensitive to this parameter.
Note that the coefficients in Eqs.~(\ref{eq:A18})
to (\ref{eq:A22}) differ slightly from those in
Ref.~\cite{RikovskaStone:2006ta}
because of the improvements in
the MIT bag model explained in Sect.~II.
The only significant change is for the
$\Sigma$ hyperon and this is a result of ensuring the
correct hyperfine splitting from the $\Lambda$.
It is worth recalling that one  underlying hypothesis of the model 
is that the $(\sigma,\omega,\rho)$ mesons do not couple to the 
strange quark, which is justified by the OZI rule since the interaction 
cannot proceed by quark exchange. Moreover, this hypothesis directly leads 
to the absence of spin-orbit splitting for $\Lambda$ hypernuclei, 
which is  consistent with experimental data. In this framework, the 
$\omega$-baryon coupling goes like  the number of non-strange quarks and 
thus follows the constraints of SU(6) symmetry. We note that, for 
example, QCD sum-rule 
investigations~\cite{Jin:1993fr}
have suggested that the $\omega$-baryon
couplings may differ from the SU(6) values. On the other hand, the success 
of the QMC model  
in leading to very realistic effective NN forces~\cite{Guichon:2006er} 
gives us some  
confidence, {\it a posteriori}, for the hypothesis. 

To calculate the hyperon levels, we use a relativistic shell model.
In principle it would be preferable to generate the
shell model core, consisting
only of nucleons, using the Hartree-Fock approximation. However, in
practice it is much easier to use Hartree approximation, which should produce
very similar results, because the coupling constants
are adjusted in each case to
reproduce the properties of nuclear matter.
Thus, we use the simpler
Hartree approximation, with free space meson
nucleon coupling constants
$g_\sigma^2 = 8.79 m_\sigma^2$, $g_\omega^2 = 4.49 m_\omega^2$
and $g_\rho^2 = 3.86 m_\rho^2$, with $m_\sigma$ = 700 MeV, $m_\omega$ =
770 MeV and $m_\rho = 780$MeV~\cite{RikovskaStone:2006ta}. Once we have
the shell model core wave functions there is, however, no practical reason
for not using the more sophisticated
Hartree-Fock couplings for the hyperon. In a  previous study of high central density
neutron stars~\cite{RikovskaStone:2006ta},
where the hyperon population
is large enough that their exchange terms matter, we found that the
Hartree-Fock couplings, $g_\sigma^2 = 11.33 m_\sigma^2$,
$g_\omega^2 = 7.27 m_\omega^2$ and $g_\rho^2 = 4.56 m_\rho^2$,
gave a satisfactory phenomenology. So, for the hyperons  we use
these couplings (in Eqs.(4-7)).

In the $1s_{1/2}$ level of $^{208}$Pb, this
yields 26.9 MeV binding for a $\Lambda$ and
a mere 3.3 MeV binding  for a $\Sigma^0$.
The reduction of more than 23 MeV binding for the $\Sigma^0$ relative to the
$\Lambda$ is a tremendous improvement over the original QMC model and
this may be traced directly to the inclusion of the effect of the medium
in enhancing the hyperfine splitting of the $\Sigma$ and $\Lambda$. 
It may be of interest to note that, if this hyperfine effect, which is an 
essential feature of the improved QMC model, were to be neglected, 
one would need, for example,  to increase the $\omega-\Sigma$ coupling 
with respect to $\omega-\Lambda$ by about 20\%.

The remarkable agreement between the calculated and
the experimental binding energy of the $\Lambda$
in the $1s_{1/2}$ level of $^{208}$Pb is another
major success. In our earlier work the $\Lambda$ was
overbound by 12 MeV and we needed to add a
phenomenological correction which we attributed to
the Pauli effect. This correction is not needed when we use Hartree-Fock, rather than Hartree, coupling
constants.
\begin{table}[htbp]
\begin{center}
\caption{Single-particle energies (in MeV)
for $^{17}_Y$O, $^{41}_Y$Ca and $^{49}_Y$Ca
hypernuclei.
The experimental data are taken from
Ref.~\protect\cite{Hashimoto:2006aw} (Table 11) for
$^{16}$O and from Ref.~\protect\cite{Pile:1991cf} for $^{40}$Ca. }
\label{spe1}
\begin{tabular}[t]{c|ccc|ccc|cc}
\hline
\hline
&$^{16}_\Lambda$O (Expt.)  &$^{17}_\Lambda$O  &$^{17}_{\Xi^0}$O
&$^{40}_\Lambda$Ca (Expt.) &$^{41}_\Lambda$Ca &$^{41}_{\Xi^0}$Ca
&$^{49}_\Lambda$Ca         &$^{49}_{\Xi^0}$Ca\\
\hline
\hline
$1s_{1/2}$&-12.42 $\pm0.05\pm0.36$      &-16.2 &-5.3 &-18.7 $\pm 1.1$       &-20.6 &-5.5 &-21.9 &-9.4 \\
$1p_{3/2}$&           & -6.4 &---  &            &-13.9 &-1.6 &-15.4 &-5.3 \\
$1p_{1/2}$& -1.85$\pm$0.06$\pm$0.36 & -6.4 &---  & &-13.9 &-1.9 &-15.4 &-5.6 \\
$1d_{5/2}$&           &      &     &            & -5.5 &---  & -7.4 &---  \\
$2s_{1/2}$&           &      &     &            & -1.0 &---  & -3.1 &---  \\
$1d_{3/2}$&           &      &     &            & -5.5 &---  & -7.3 &---  \\
\end{tabular}
\end{center}
\end{table}

Already at this stage the binding of the $\Sigma^0$ in the $1s_{1/2}$
level of $^{208}$Pb is just a few MeV -- a major improvement over the earlier
QMC results. However, as pointed out in that work, there is an additonal
piece of physics which really should be included
and which goes beyond the
naive description of the intermediate range attraction in terms of
$\sigma$ exchange. In particular, the energy
released in the
two-pion exchange process, N $\Sigma \ra$ N $\Lambda \ra$ N $\Sigma$,
because of the $\Sigma$--$\Lambda$ mass difference,
reduces the intermediate range attraction felt by the $\Sigma$ hyperon.
In Ref.~\cite{Tsushima:1997cu} this was modeled
by introducing an additional
vector repulsion for a $\Sigma$ hyperon. Following the same
procedure, we replace $g_\omega^{\Sigma} \omega(r)$
by $g_\omega^{\Sigma} \omega(r)
+ \lambda_\Sigma \rho_B$,
with $\lambda_\Sigma = 50.3$ MeV-fm$^3$,
as determined
in Ref.~\cite{Tsushima:1997cu} by comparison with
the more microscopic model of the
J\"ulich group~\cite{Reuber:1993ip}. We are comfortable using this earlier
estimate of the effect of the coupled $\Lambda-N$ channel, even though
there is a more recent potential from the J\"ulich
group~\cite{Haidenbauer:2005zh}, because the latter tends to overbind
hypernuclei~\cite{Sammarruca:2008hy} and the sign of the correction for
coupling to a lower mass channel is in any case model independent.
\begin{table}[htbp]
\begin{center}
\caption{Same as table~\ref{spe1} but
for $^{91}_Y$Zr and $^{208}_Y$Pb hypernuclei. The experimental data are
taken from Ref.~\protect\cite{Hashimoto:2006aw} (Table 13).}
\label{spep2}
\begin{tabular}[t]{c|ccc|ccc}
\hline
\hline
&$^{89}_\Lambda$Yb (Expt.)  &$^{91}_\Lambda$Zr  &$^{91}_{\Xi^0}$Zr
&$^{208}_\Lambda$Pb (Expt.) &$^{209}_\Lambda$Pb &$^{209}_{\Xi^0}$Pb \\
\hline
\hline
$1s_{1/2}$&-23.1 $\pm 0.5$       &-24.0 &-9.9 &-26.3 $\pm 0.8$&-26.9 &-15.0 \\
$1p_{3/2}$&            &-19.4 &-7.0 &            &-24.0 &-12.6 \\
$1p_{1/2}$&-16.5 $\pm 4.1$ ($1p$)&-19.4 &-7.2 &-21.9 $\pm 0.6$ ($1p$)&-24.0 &-12.7 \\
$1d_{5/2}$&            &-13.4 &-3.1 &---         &-20.1 & -9.6 \\
$2s_{1/2}$&            & -9.1 &---  &---         &-17.1 & -8.2 \\
$1d_{3/2}$&-9.1 $\pm 1.3$  ($1d$)&-13.4 &-3.4 &-16.8 $\pm 0.7$ ($1d$)&-20.1 & -9.8 \\
$1f_{7/2}$&            & -6.5 &---  &---         &-15.4 & -6.2 \\
$2p_{3/2}$&            & -1.7 &---  &---         &-11.4 & -4.2 \\
$1f_{5/2}$&-2.3 $\pm1.2$  ($1f$)& -6.4 &---  &-11.7 $\pm 0.6$ ($1f$)&-15.4 & -6.5 \\
$2p_{1/2}$&            & -1.6 &---  &---         &-11.4 & -4.3 \\
$1g_{9/2}$&            &---   &---  &---         &-10.1 & -2.3 \\
$1g_{7/2}$&            &---   &---  & -6.6 $\pm 0.6$ ($1g$)&-10.1 & -2.7 \\
$1h_{11/2}$&           &---   &---  &---         & -4.3 &---   \\
$2d_{5/2}$&            &---   &---  &---         & -5.3 &---   \\
$2d_{3/2}$&            &---   &---  &---         & -5.3 &---   \\
$1h_{9/2}$&            &---   &---  &---         & -4.3 &---   \\
$3s_{1/2}$&            &---   &---  &---         & -3.5 &---   \\
\end{tabular}
\end{center}
\end{table}

Our results are presented in Tables~\ref{spe1} and \ref{spep2}.
The overall
agreement with the experimental energy levels of $\Lambda$ hypernuclei
across the periodic table is quite good. The discrepancies
which remain may well be
resolved by small effective hyperon-nucleon interactions which go beyond
the simple, single-particle shell model. Once again, we stress the very
small spin-orbit force experienced by the $\Lambda$, which is a natural
property of the QMC model~\cite{Tsushima:1997rd}.

\begin{figure}[t]
\centerline{\includegraphics[clip,angle=0,width=9cm]{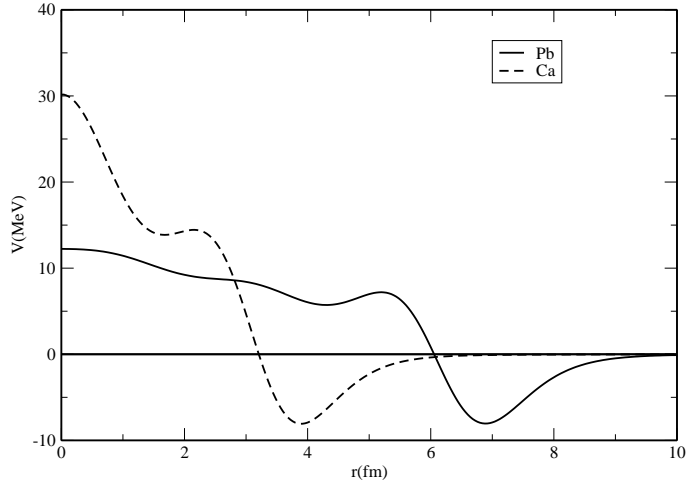}}
\caption{\label{fig:pot} The effective  potential for $\Sigma^{0}$ in Calcium and Lead}
\end{figure}

There are no entries for the
$\Sigma$-hyperon because neither the $\Sigma^+$
nor the $\Sigma^0$ is bound to a finite nucleus.
This absence of bound
$\Sigma$-hypernuclei constitutes a major advance over
earlier work. We stress that this is a direct
consequence of the enhancement of the hyperfine
interaction (that splits the masses of the
$\Sigma$ and $\Lambda$ hyperons) by the
mean scalar field in-medium. It is especially
interesting to examine the effective
non-relativistic potential felt by the $\Sigma^0$
in a finite nucleus. This is shown in Fig.\ref{fig:pot} for Calcium and Lead.  In the central region the vector interaction  dominates over the scalar one  leading to  a repulsive effective potential which reaches respectively 30 MeV and 12 MeV at the center. It is only at the surface that the scalar attraction   becomes dominant.
While the exact numerical values depend on the
mass taken for the $\sigma$ meson, we stress
the similarity to
the phenomenological form found by
Batty {\it et al.}~\cite{Batty:1994sw}. For a recent review see \cite{Friedman:2007qx}. It will clearly be very interesting to pursue the
application of the current theoretical formulation
to $\Sigma^-$-atoms.

We also note that this model supports the
existence of a variety of
bound $\Xi$-hypernuclei. For the $\Xi^0$ the binding
of the 1s level varies from 5 MeV in
$^{17}_{\Xi^0}$O to
15 MeV in $^{209}_{\Xi^0}$Pb. The experimental
search for such states at facilities such as
J-PARC and GSI-FAIR will be very important.

\section{Concluding remarks}

In conclusion, we stress that including the
effect of the medium on the hyperfine interaction
between quarks within the quark-meson coupling
model has led to some important advances. The
agreement between the parameter free
calculations and the experimental ground state  levels
for  $\Lambda$-hypernuclei from Calcium to Lead 
is impressive. In Oxygen the agreement is not as good and we take 
it as a reminder that the QMC model may be better at describing heavy nuclei,  
for which the constant density region dominates over the surface. For 
the d- and f-wave levels
shown in Table III, there is a tendency for the
model to overbind by several MeV. Whether this
is a consequence of the use of an extreme single
particle shell model for the core, the omission
of residual $\Lambda-N$ interactions or an aspect
of the current implementation of QMC that requires
improvement remains to be seen. Nevertheless,
we find these initial results, obtained with no
adjustment of parameters to hyperon data, very
encouraging.

A number of $\Xi$-hypernuclei are predicted to be
bound, although not as deeply as in the $\Lambda$
case. On the other hand, the additional repulsion
arising from the enhancement of the hyperfine
reulsion in the $\Sigma$-hyperon in-medium,
together with the effect of the $\Sigma N -
\Lambda N$ channel coupling on the intermediate
range scalar attraction, means that
no $\Sigma$-hypernuclei are predicted to be bound.
This encouraging picture of finite
hypernuclei, suggests that the underlying model, which
is fully relativistic and incorporates the
quark substructure of the baryons, is ideally
suited for application to the properties of
dense matter and neutron stars.

\begin{acknowledgments}
This work was supported by the Espace de Structure
Nucl\'eaire Th\'eorique du CEA and in part by DOE contract DE-AC05-06OR23177,
under which Jefferson Science Associates, LLC,
operates Jefferson Lab.
K.T. was supported by the Spanish Ministry of Education and Science,
Reference Number: SAB2005-0059 in the initial stage of this work.
\end{acknowledgments}
%

%

\begin{thebibliography}{99}
%
\bibitem{Dalitz:1978hw}
  R.~H.~Dalitz and A.~Gal,
  Annals Phys.\  {\bf 116} (1978) 167.
%
\bibitem{Povh:1987pt}
B.~Povh,
Prog.\ Part.\ Nucl.\ Phys.\  {\bf 18} (1987) 183.
%
\bibitem{Yamamoto:1994tc}
Y.~Yamamoto, T.~Motoba, H.~Himeno, K.~Ikeda and S.~Nagata,
Prog.\ Theor.\ Phys.\ Suppl.\  {\bf 117} (1994) 361.
%
\bibitem{Gal:2004cf}
A.~Gal,
Prog.\ Theor.\ Phys.\ Suppl.\  {\bf 156} (2004) 1.
%
\bibitem{Hashimoto:2006aw}
  O.~Hashimoto and H.~Tamura,
  Prog.\ Part.\ Nucl.\ Phys.\  {\bf 57} (2006) 564.
%
\bibitem{Ajimura:2001na}
S.~Ajimura {\it et al.},
  Phys.\ Rev.\ Lett.\  {\bf 86} (2001) 4255.
%
\bibitem{JLab-Expt}
JLab Proposal E01-011 (2001),
Spokespersons O.~Hashimoto, L.~Tang, J.~Reinhold and S.~N.~Nakamura.
%
\bibitem{Bart:1999uh}
S.~Bart {\it et al.},
Phys.\ Rev.\ Lett.\  {\bf 83} (1999) 5238.
%
\bibitem{Noumi:2001tx}
H.~Noumi {\it et al.},
Phys.\ Rev.\ Lett.\  {\bf 89} (2002) 072301
  [Erratum-ibid.\  {\bf 90} (2003) 049902].
%
\bibitem{Saha:2004ha}
  P.~K.~Saha {\it et al.},
  Phys.\ Rev.\  C {\bf 70} (2004) 044613.
%
\bibitem{Fukuda:1998bi}
  T.~Fukuda {\it et al.}  [E224 Collaboration],
  Phys.\ Rev.\  C {\bf 58} (1998) 1306.
%
\bibitem{Khaustov:1999bz}
  P.~Khaustov {\it et al.}  [AGS E885 Collaboration],
  Phys.\ Rev.\  C {\bf 61} (2000) 054603.
%
\bibitem{Weber:2004kj}
F.~Weber,
Prog.\ Part.\ Nucl.\ Phys.\  {\bf 54} (2005) 193.
%
\bibitem{SchaffnerBielich:2004ch}
J.~Schaffner-Bielich,
J.\ Phys.\ G {\bf 31} (2005) S651.
%
\bibitem{RikovskaStone:2006ta}
  J.~Rikovska-Stone, P.~A.~M.~Guichon, H.~H.~Matevosyan and A.~W.~Thomas,
  Nucl.\ Phys.\  A {\bf 792} (2007) 341.
%
\bibitem{Jin:1993fr}
X.~M.~Jin and R.~J.~Furnstahl,
Phys.\ Rev.\  C {\bf 49} (1994) 1190.
%
\bibitem{Guichon:2006er}
P.~A.~M.~Guichon, H.~H.~Matevosyan, N.~Sandulescu and A.~W.~Thomas,
Nucl.\ Phys.\  A {\bf 772} (2006) 1.
%
\bibitem{Guichon:1987jp}
P.~A.~M.~Guichon,
  Phys.\ Lett.\  B {\bf 200} (1988) 235.
%
\bibitem{Guichon:1995ue}
P.~A.~M.~Guichon, K.~Saito, E.~N.~Rodionov and A.~W.~Thomas,
  Nucl.\ Phys.\  A {\bf 601} (1996) 349.
%
\bibitem{Saito:1994kg}
K.~Saito and A.~W.~Thomas,
Phys.\ Rev.\  C {\bf 51} (1995) 2757.
%
\bibitem{Tsushima:1998qw}
K.~Tsushima, D.~H.~Lu, A.~W.~Thomas and K.~Saito,
Phys.\ Lett.\  B {\bf 443} (1998) 26.
%
\bibitem{Bass:2005hn}
S.~D.~Bass and A.~W.~Thomas,
Phys.\ Lett.\  B {\bf 634} (2006) 368.
%
\bibitem{Tsushima:1998ru}
  K.~Tsushima, D.~H.~Lu, A.~W.~Thomas, K.~Saito and R.~H.~Landau,
  Phys.\ Rev.\  C {\bf 59} (1999) 2824.
\bibitem{Saito:2005rv}
K.~Saito, K.~Tsushima and A.~W.~Thomas,
Prog.\ Part.\ Nucl.\ Phys.\  {\bf 58} (2007) 1.
%
\bibitem{Tsushima:1997cu}
K.~Tsushima, K.~Saito, J.~Haidenbauer and A.~W.~Thomas,
Nucl.\ Phys.\  A {\bf 630} (1998) 691.
%
\bibitem{Tsushima:1997rd}
K.~Tsushima, K.~Saito and A.~W.~Thomas,
Phys.\ Lett.\  B {\bf 411} (1997) 9
[Erratum-ibid.\  B {\bf 421} (1998) 413].
%
\bibitem{Bentz:2001vc}
W.~Bentz and A.~W.~Thomas,
  Nucl.\ Phys.\  A {\bf 696} (2001) 138.
%
\bibitem{Cloet:2005rt}
I.~C.~Cloet, W.~Bentz and A.~W.~Thomas,
Phys.\ Rev.\ Lett.\  {\bf 95} (2005) 052302.
%
\bibitem{Cloet:2006bq}
I.~C.~Cloet, W.~Bentz and A.~W.~Thomas,
Phys.\ Lett.\  B {\bf 642} (2006) 210.
%
\bibitem{Chodos:1974pn}
A.~Chodos, R.~L.~Jaffe, K.~Johnson and C.~B.~Thorn,
Phys.\ Rev.\  D {\bf 10} (1974) 2599.
%
\bibitem{Barnes:1984pw}
T.~Barnes,
Phys.\ Rev.\  D {\bf 30} (1984) 1961.
%
\bibitem{Reuber:1993ip}
A.~Reuber, K.~Holinde and J.~Speth,
Nucl.\ Phys.\  A {\bf 570} (1994) 543.
%
\bibitem{Pile:1991cf}
P.~H.~Pile {\it et al.},
Phys.\ Rev.\ Lett.\  {\bf 66} (1991) 2585.
%
\bibitem{Haidenbauer:2005zh}
J.~Haidenbauer and U.~G.~Meissner,
Phys.\ Rev.\  C {\bf 72} (2005) 044005.
%
\bibitem{Sammarruca:2008hy}
F.~Sammarruca,
``Predicting the Lambda binding energy in nuclear matter,''
arXiv:0801.0879 [nucl-th].
%
\bibitem{Batty:1994sw}
C.~J.~Batty, E.~Friedman and A.~Gal,
Prog.\ Theor.\ Phys.\ Suppl.\  {\bf 117} (1994) 227.
%
\bibitem{Friedman:2007qx}
E.~Friedman and A.~Gal,
Phys.\ Rep.\ {\bf 452} (2007) 89.
\end{thebibliography}
\end{document}